\documentclass[a4paper,11pt]{article}

\usepackage{jheppub}
\usepackage{amsfonts,graphics,epsfig,subfigure}
\usepackage{color,graphicx}
\usepackage{amsfonts,amssymb,theorem,mathrsfs,times}
\usepackage{bm}

\title{Coexistence curve and molecule number density of AdS topological charged black hole in massive gravity}

\author[a]{Yi-Fei Wang}
\author[a]{Ming Zhang}
\author[a,1]{Wen-Biao Liu,\note{Corresponding author}}

\affiliation[a]{Department of Physics, Institute of Theoretical Physics,  Beijing Normal University, Beijing, 100875, China}

\emailAdd{201521140018@mail.bnu.edu.cn}
\emailAdd{201421140015@mail.bnu.edu.cn}
\emailAdd{wbliu@bnu.edu.cn}

\abstract{The coexistence curve and molecule number density of a $4$-dimensional AdS topological charged black hole in massive gravity is investigated. We find that the analytic expression of the coexistence curve in the reduced parameter space is dependent on theory parameters. This is very different from the previous results obtained in other modified gravity such as $f(R)$ gravity and Gauss-Bonnet gravity. Besides, we derive the explicit expression of the physical quantity which describes the difference of the number densities of AdS topological charged black hole molecules between the small and large black hole. It is observed that the difference of the molecule number densities is also dependent on theory parameters. Both the expressions of the coexistence curve and the difference of the molecule number densities can be reduced into a form which is similar to a RN-AdS black hole if the mass of graviton $m$ is zero. Moreover, we find the shifted temperature under massive gravity. This can highlight the important role played by the mass of graviton and other parameters in the phase transitions of AdS black holes in massive gravity.}

\begin{document}
\maketitle
\flushbottom

\section{Introduction}\label{Sec1}

The pioneering work by Hawking and Page\cite{Hawking:1982dh} in $1983$ triggered many subsequent researches concerning AdS black hole thermodynamics. Asymptotically AdS black holes have attracted much attention for the advent of AdS/CFT correspondence\cite{Maldacena:1997re,Gubser:1998bc,Witten:1998qj}, which says that a quantum gravity in AdS space is dual to a conformal field theory (CFT) living on the boundary of the AdS space. When we consider charged AdS black holes, there exists a first--order phase transition between small black holes and large black holes if the charge is smaller than a critical value\cite{Chamblin:1999tk,Chamblin:1999hg,Banerjee:2011au,Banerjee:2011raa}. As the temperature increases to the critical value, the coexistence curves for phase transitions of black holes have critical points where the phase transitions terminate and the first--order phase transitions become second--order ones. Researchers established an analogy between phase transitions of classical charged AdS black holes and phase transitions of Van der Waals liquid-gas system. The complete analogy was constructed when physicists realized that the cosmological constant should be treated as a pressure ($P=-\Lambda/8\pi$) and its conjugate quantity as the thermodynamic volume\cite{Dolan:2011xt,Cvetic:2010jb,Dolan:2013ft,Kastor:2010gq,Castro:2013pqa,El-Menoufi:2013pza,Kubiznak:2012wp}. There also exists much more interesting phenomena besides the VdW-type first-order phase transition\cite{Gunasekaran:2012dq,Altamirano:2013ane,Wei:2014hba,Frassino:2014pha,Altamirano:2013uqa}. Such investigations revealed much richer physics of thermodynamics of black holes and gave some enlightenment for the following researches.

Recently, S. Wei \emph{et al.} suggested an insight into the microscopic structure of RN-AdS black holes based on its thermodynamical phase transition in the extended phase space\cite{Wei:2015iwa}. They proposed an important quantity to describe the microscopic degrees of freedom of the black hole in the phase transition: number density $n$. They found that the number density suffers a sudden change accompanied by a latent heat when the black hole system goes across the SBH/LBH coexistent curve. When the system passes the critical point, it encounters a second-order phase transition with a vanishing latent heat due to the continuous change of the number density. Their work was generalized by J. Mo \emph{et al.} to the case of $f(R)$ black holes and Gauss-Bonnet AdS black holes in modified gravity\cite{Mo:2016sel}. It is found that both the coexistence curve and the difference of the number densities of a charged $f(R)$ AdS black hole coincide with those of a RN-AdS black hole while an uncharged Gauss-Bonnet AdS black hole behaves differently. This is because the non-trivial charge has effects on the lower limit of the inner horizon radius of the extremal black hole.

General relativity established by Einstein is a relativistic theory of gravity where the graviton is massless. A natural question is whether one can build a self-consistent gravity theory if the graviton is massive. The possibility of a massive graviton has been studied firstly by Fierz and Pauli\cite{Fierz:1939ix}. A class of nonlinear massive gravity theories were proposed in Refs.\cite{deRham:2010ik,deRham:2010kj,Hinterbichler:2011tt}, in which the ghost field is absent\cite{Hassan:2011hr,Hassan:2011tf}. There are some papers concerning the corresponding thermodynamical properties and phase structure of black holes in massive gravity\cite{Cai:2014znn,Xu:2015rfa}. On the other hand, Van der Waals behavior and its applications in Einstein gravity are found only for AdS black holes with spherical horizon topology\cite{Dolan:2010ha,Kubiznak:2012wp}. However, it is shown in recent investigations that black holes in massive gravity exhibit Van der Waals behavior independent on the choice of horizon curvature\cite{Xu:2015rfa,Hendi:2017fxp}. It motivates one to explore what effects the curvature and the parameters in massive gravity may have to the coexistence curve and difference of number densities of AdS black hole phase transitions.

The organization of this paper is as follows. In Sec.\ref{Sec2}, we will briefly review the black hole solution and its corresponding thermodynamical properties, then the equation of state in reduced parameter space will be given. The explicit expression of coexistence curve and the difference of molecule number density will be investigated in Sec.\ref{Sec3}. Sec.\ref{Sec4} will be dedicated to discussions and conclusions.

\section{The black hole solution and the equation of state in reduced parameter space}\label{Sec2}

Let us consider the following action for an ($n+2$)-dimensional massive gravity with a Maxwell field and a negative cosmological constant\cite{Xu:2015rfa}

\begin{equation}
S=\frac{1}{16\pi}\int
d^{n+2}x\sqrt{-g}\Big[R+\frac{n(n+1)}{l^2}-\frac{1}{4}F^2+m^2\sum_{i=1}^4c_i\mathcal{U}_i(g,f)\Big],
\end{equation}
where $R$ is the scalar curvature of the metric $g$, $F$ is the Maxwell invariant, $m$ is the mass of graviton, $c_i$ are constants, $f$ is a fixed symmetric tensor called the reference metric, and $\mathcal{U}_i$ are symmetric polynomials of eigenvalues of the ($n+2$)$\times$($n+2$) matrix
$\mathcal{K}^{\mu}_{~\nu}\equiv\sqrt{g^{\mu\alpha}f_{\alpha\nu}}$, which read

\begin{eqnarray}
&~&\mathcal{U}_1=[\mathcal{K}],\nonumber\\
&~&\mathcal{U}_2=[\mathcal{K}]^2-[\mathcal{K}^2],\nonumber\\
&~&\mathcal{U}_3=[\mathcal{K}]^3-3[\mathcal{K}][\mathcal{K}^2]+2[\mathcal{K}^3],\nonumber\\
&~&\mathcal{U}_4=[\mathcal{K}]^4-6[\mathcal{K}^2][\mathcal{K}]^2+8[\mathcal{K}^3][\mathcal{K}]+3[\mathcal{K}^2]^2-6[\mathcal{K}^4].
\end{eqnarray}
The square root in $\mathcal{K}$ can be interpreted as a matrix square root, \emph{i.e.},
$(\sqrt{A})^{\mu}_{~\nu}(\sqrt{A})^{\nu}_{~\lambda}=A^{\mu}_{~\lambda}$,
and the traces are expressed as
$[\mathcal{K}]=\mathcal{K}^{\mu}_{~\mu}$ by the rectangular brackets.

In order to obtain AdS topological static charged black hole in massive gravity, we consider the metric of an ($n+2$)--dimensional spacetime as

\begin{equation}
ds^2=-f(r)dt^2+f^{-1}(r)dr^2+r^2h_{ij}dx^idx^j.
\end{equation}
Accordingly, we employ the ansatz as\cite{Xu:2015rfa}

\begin{equation}\label{refmetric}
f_{\mu\nu}=\mathrm{diag}(0,0,c_0^2h_{ij}),
\end{equation}
where $c_0$ is a positive constant, and $h_{ij}dx^idx^j$ expresses the line element for an Einstein space with constant curvature $n(n-1)k$. One can set $k$ to be $1$, $0$, or $-1$ without loss of generality, corresponding to a spherical, Ricci flat, or hyperbolic topology of the black hole horizon, respectively. According to the reference metric (\ref{refmetric}), we have

\begin{eqnarray}\label{utermeincoor}
&~&\mathcal{U}_1=nc_0/r,\nonumber\\
&~&\mathcal{U}_2=n(n-1)c_0^2/r^2,\nonumber\\
&~&\mathcal{U}_3=n(n-1)(n-2)c_0^3/r^3,\nonumber\\
&~&\mathcal{U}_4=n(n-1)(n-2)(n-3)c_0^4/r^4.
\end{eqnarray}
The metric function $f(r)$ is then given by\cite{Cai:2014znn}

\begin{eqnarray}\label{f}
f(r)&=&k+\frac{16\pi P}{(n+1)n}r^2-\frac{16\pi
M}{nV_nr^{n-1}}+\frac{(16\pi Q)^2}{2n(n-1)V_n^2r^{2(n-1)}}+\frac{c_0c_1m^2}{n}r+c_0^2c_2m^2\nonumber\\
&~&+\frac{(n-1)c_0^3c_3m^2}{r}+\frac{(n-1)(n-2)c_0^4c_4m^2}{r^2},
\end{eqnarray}
where $V_n$ denotes the volume of space spanned by coordinates $x^i$, $M$ is the mass of black hole, $Q$ is a constant in terms of the black hole charge, and $P=\frac{n(n+1)}{16\pi l^2}$ is the pressure. The black hole horizon is determined by $f(r)|_{r=r_h}=0$. Therefore, we can express the mass $M$ in terms of $r_h$ as\cite{Xu:2015rfa}

\begin{eqnarray}\label{enthalpy}
M&=&\frac{nV_nr_h^{n-1}}{16\pi}\Big[k+\frac{16\pi
P}{(n+1)n}r_h^2+\frac{(16\pi Q)^2}{2n(n-1)V_n^2r_h^{2(n-1)}}+\frac{c_0c_1m^2}{n}r_h+c_0^2c_2m^2\nonumber\\
&~&+\frac{(n-1)c_0^3c_3m^2}{r_h}+\frac{(n-1)(n-2)c_0^4c_4m^2}{r_h^2}\Big].
\end{eqnarray}
It is not difficult to get the Hawking temperature by removing the conical singularity at the horizon in the Euclidean sector of the black hole solution, which is given by\cite{Xu:2015rfa}

\begin{eqnarray}\label{Htemp}
T=\frac{1}{4\pi}f'(r_h)&=&\frac{1}{4\pi r_h}\Big[(n-1)k+\frac{16\pi
P}{n}r_h^2-\frac{(16\pi Q)^2}{2nV_n^2r_h^{2(n-1)}}+c_0c_1m^2r_h+(n-1)c_0^2c_2m^2\nonumber\\
&~&+\frac{(n-1)(n-2)c_0^3c_3m^2}{r_h}+\frac{(n-1)(n-2)(n-3)c_0^4c_4m^2}{r_h^2}\Big].
\end{eqnarray}

The expression of entropy is given by

\begin{equation}\label{entropy}
S=\int_0^{r_h}T^{-1}\left(\frac{\partial H}{\partial
r}\right)_{Q,P}dr=\frac{V_n}{4}r_h^n.
\end{equation}

The Gibbs free energy is defined as

\begin{equation}\label{Gibbs}
G=H-TS=M-TS.
\end{equation}
Note that the mass $M$ should be understood as the enthalpy instead of internal energy in the extended phase space and in the above derivation we have utilized the identification $H\equiv M$.

Now we focus on the $4$-dimensional spacetime case ($n=2$). From the Hawking temperature (\ref{Htemp}) in the $n=2$ case we can obtain the black hole's equation of state as\cite{Xu:2015rfa}

\begin{equation}\label{P}
P=\Big(\frac{T}{2}-\frac{c_0c_1m^2}{8\pi}\Big)\frac{1}{r_h}-\Big(\frac{k}{8\pi}+\frac{c_0^2c_2m^2}{8\pi}\Big)\frac{1}{r_h^2}+\frac{8\pi
Q^2}{V_2^2}\frac{1}{r_h^4}.
\end{equation}
Now we need to introduce the specific volume for the black hole which is defined as\cite{Wei:2015iwa}

\begin{equation}\label{v}
v={2l_P^2r_h},
\end{equation}
where we restore the dimension and Planck length $l_P=\sqrt{\hbar G/c^3}$.
Now the black hole's phase structure can be studied in the canonical ensemble whose charge is fixed in terms of $P$-$v$ diagram. Note that the specific volume $v$ as Eq.(\ref{v}) is a monotonic function of the horizon radius $r_h$, so we can use $r_h$ rather than $v$ to specify the critical behavior. By the condition of the inflection point in the $P$-$v$ diagram we can determine the critical point as

\begin{equation}\label{cricondition}
\frac{\partial P}{\partial
r_h}\Big|_{r_h=r_{hc},T=T_c}=\frac{\partial^2P}{\partial
r_h^{~2}}\Big|_{r_h=r_{hc},T=T_c}=0.
\end{equation}
Therefore, we can figure out the critical quantities of the critical point which is determined by Eq.(\ref{cricondition}) as
\begin{small}
\begin{equation}\label{criquan}
v_{c}\!=\!2r_{hc}\!=\!\frac{16\sqrt{6}\pi Q}{\sqrt{k+c_0^2c_2m^2}V_2}, P_c\!=\!\frac{(k+c_0^2c_2m^2)^2V_2^2}{6144\pi^3 Q^2}, T_c\!=\!\frac{(k+c_0^2c_2m^2)^{3/2}V_2+6\sqrt6\pi Qc_0c_1m^2}{24\sqrt{6}\pi^2 Q}.
\end{equation}
\end{small}
To get the equation of state in the reduced parameter space we shall introduce the following definitions

\begin{equation}\label{reducedquan}
p=\frac{P}{P_c}, \tau=\frac{T}{T_c}, \nu=\frac{v}{v_c}.
\end{equation}
And the equation of state (\ref{P}) can thus be cast into a reduced form as follws

\begin{eqnarray}\label{eosreduced}
&\tau=\frac{6\sqrt6\pi Qc_0c_1m^2}{(k+c_0^2c_2m^2)^{3/2}V_2+6\sqrt6\pi Qc_0c_1m^2}+\frac{3\sqrt6(k+c_0^2c_2m^2)^2V_2}{8\sqrt6(k+c_0^2c_2m^2)^2V_2+288\pi Qc_0c_1m^2(k+c_0^2c_2m^2)^{1/2}}p\nu\nonumber\\
&+\frac{3(k+c_0^2c_2m^2)^{3/2}V_2}{4(k+c_0^2c_2m^2)^{3/2}V_2+24\sqrt6\pi Qc_0c_1m^2}\frac{1}{\nu}-\frac{\pi Q^2(k+c_0^2c_2m^2)^{3/2}V_2}{8\pi Q^2(k+c_0^2c_2m^2)^{3/2}V_2+48\sqrt{6}\pi^2 Q^3c_0c_1m^2}\frac{1}{\nu^3}.
\end{eqnarray}

\section{The coexistence curve and the difference of molecule number density for AdS topological charged black hole}\label{Sec3}

Now we are going to study coexistence curve in the reduced parameter space. Substituting Eqs.(\ref{enthalpy}), (\ref{Htemp}), (\ref{entropy}) into Eq.(\ref{Gibbs}) and utilizing Eq.(\ref{v}), one can obtain the explicit expression of Gibbs free energy for $4$-dimensional case as follow

\begin{equation}\label{Gibbs2}
G=\frac{24\pi Q^2}{V_2}\frac{1}{v}+\frac{(k+c_0^2c_2m^2)V_2}{32\pi}v-\frac{PV_2}{48}v^3.
\end{equation}
Substituting Eq.(\ref{criquan}) into Eq.(\ref{Gibbs2}), we obtain

\begin{equation}\label{Gibbscri}
G_c=\frac{2\sqrt{6}(k+c_0^2c_2m^2)^{1/2}Q}{3},
\end{equation}
where $G_c$ signifies the Gibbs free energy at the critical point.
In order to investigate the behavior of Gibbs free energy in the reduced parameter space, the reduced Gibbs free energy shall be introduced as

\begin{equation}\label{reducedGibbs}
\tilde{G}=\frac{G}{G_c}.
\end{equation}
Utilizing Eqs.(\ref{criquan}), (\ref{reducedquan}), (\ref{Gibbs2}), (\ref{Gibbscri}) and (\ref{reducedGibbs}), one can obtain

\begin{equation}\label{reducedGibbs2}
\tilde{G}=\frac{3}{8\nu}+\frac{3}{4}\nu-\frac{1}{8}p\nu^3.
\end{equation}
We can see clearly from the above equation that the reduced Gibbs free energy is independent on the parameters associated with massive gravity.

\begin{figure*}
\centerline{\subfigure[]{\label{1a}
\includegraphics[width=8cm,height=6cm]{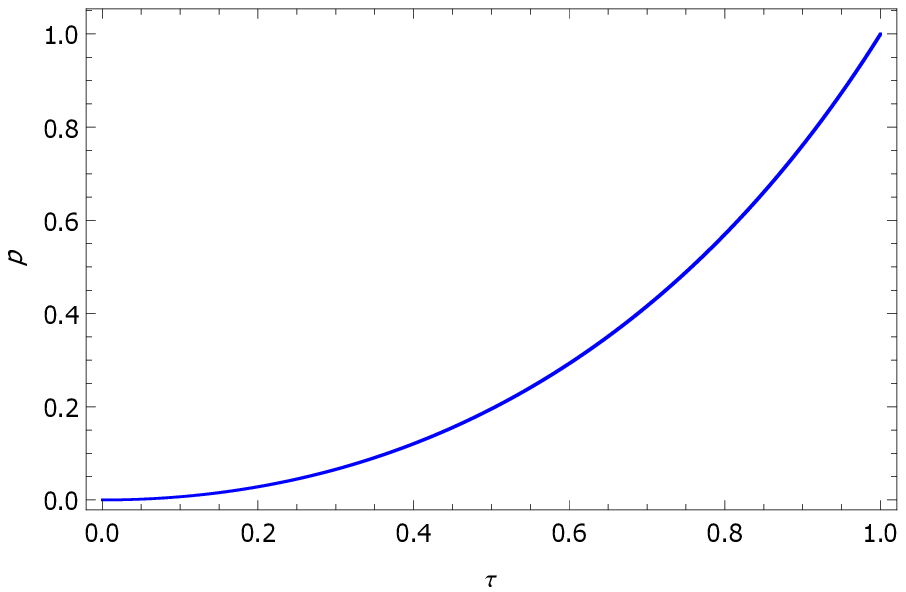}}
\subfigure[]{\label{1b}
\includegraphics[width=8cm,height=6cm]{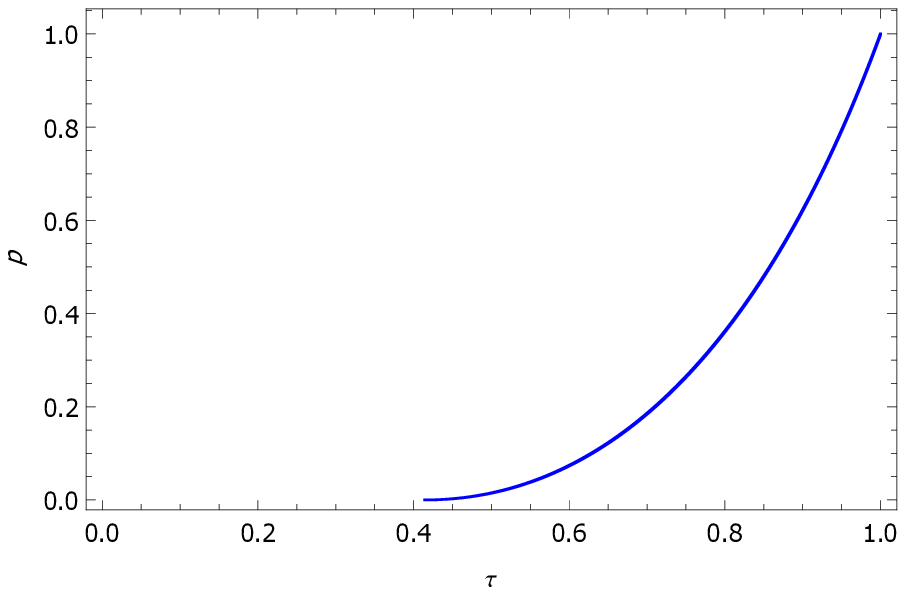}}}
 \caption{(a) coexistence curve of massive gravity AdS black hole in the reduced parameter space when $\alpha=0$ (b) coexistence curve of massive gravity AdS black hole in the reduced parameter space when $\alpha=1$} \label{fg1}
\end{figure*}

Now we begin to study two phases living at an arbitrary point of the coexistence curve. The reduced physical quantities of the point which we studied are denoted as $\nu_1,\;p,\;\tau,\;\tilde{G}_1$ and $\nu_2,\;p,\;\tau,\;\tilde{G}_2$, respectively. According to Eqs.(\ref{reducedGibbs2}) and (\ref{eosreduced}), it is not difficult to obtain the following equations
\begin{eqnarray}
\tilde{G}_1&=&\frac{3+6\nu_1^2-p\nu_1^4}{8\nu_1},\label{tildeG1}
\\
\tilde{G}_2&=&\frac{3+6\nu_2^2-p\nu_2^4}{8\nu_2},\label{tildeG2}
\\
\tau&=&C_1+C_2p\nu_1+C_3\frac{1}{\nu_1}-C_4\frac{1}{\nu_1^3},\label{tau1}
\\
\tau&=&C_1+C_2p\nu_2+C_3\frac{1}{\nu_2}-C_4\frac{1}{\nu_2^3},\label{tau2}
\end{eqnarray}
where
\begin{equation}
C_1=\frac{6\sqrt6\pi Qc_0c_1m^2}{(k+c_0^2c_2m^2)^{3/2}V_2+6\sqrt6\pi Qc_0c_1m^2},\nonumber
\end{equation}
\begin{equation}
C_2=\frac{3\sqrt6(k+c_0^2c_2m^2)^2V_2}{8\sqrt6(k+c_0^2c_2m^2)^2V_2+288\pi Qc_0c_1m^2(k+c_0^2c_2m^2)^{1/2}},\nonumber
\end{equation}
\begin{equation}
C_3=\frac{3(k+c_0^2c_2m^2)^{3/2}V_2}{4(k+c_0^2c_2m^2)^{3/2}V_2+24\sqrt6\pi Qc_0c_1m^2},\nonumber
\end{equation}
\begin{equation}
C_4=\frac{\pi Q^2(k+c_0^2c_2m^2)^{3/2}V_2}{8\pi Q^2(k+c_0^2c_2m^2)^{3/2}V_2+48\sqrt{6}\pi^2 Q^3c_0c_1m^2}.\nonumber
\end{equation}
The points on the coexistence curve represent first order phase transition of the black holes. So the two phases hold identical temperature and Gibbs free energy. As a result, Eqs.(\ref{tildeG1})-(\ref{tau2}) can be reorganized as

\begin{equation}
\frac{3+6\nu_1^2-p\nu_1^4}{8\nu_1}=\frac{3+6\nu_2^2-p\nu_2^4}{8\nu_2},\label{core1}
\end{equation}
\begin{equation}
C_1+C_2p\nu_1+C_3\frac{1}{\nu_1}-C_4\frac{1}{\nu_1^3}=C_1+C_2p\nu_2+C_3\frac{1}{\nu_2}-C_4\frac{1}{\nu_2^3},\label{core2}
\end{equation}
\begin{equation}
2\tau=C_1+C_2p\nu_1+C_3\frac{1}{\nu_1}-C_4\frac{1}{\nu_1^3}+C_1+C_2p\nu_2+C_3\frac{1}{\nu_2}-C_4\frac{1}{\nu_2^3}.\label{core3}
\end{equation}
Solving these equations, one can obtain
\begin{eqnarray}
p=3&-\frac{(-1)^{1/3}\Big[54+\sqrt{(\alpha\tau+\sqrt{2}\tau-\alpha)^6[(\alpha\tau+\sqrt{2}\tau-\alpha)^2-4]}+(\alpha\tau+\sqrt{2}\tau-\alpha)^4-18(\alpha\tau+\sqrt{2}\tau-\alpha)^2\Big]^{1/3}}{2^{1/3}}\nonumber\\
&+\frac{(-1)^{2/3}2^{1/3}[9-2(\alpha\tau+\sqrt{2}\tau-\alpha)^2]}
{\Big[54+\sqrt{(\alpha\tau+\sqrt{2}\tau-\alpha)^6[(\alpha\tau+\sqrt{2}\tau-\alpha)^2-4]}+(\alpha\tau+\sqrt{2}\tau-\alpha)^4-18(\alpha\tau+\sqrt{2}\tau-\alpha)^2\Big]^{1/3}},\label{cocur}
\end{eqnarray}
where $\alpha=\frac{12\sqrt{3}c_0c_1m^2\pi Q}{(k+c_0^2c_2m^2)^{3/2}V_2}$.
If we ignore the parameters accociated with massive gravity, \emph{i.e.}, setting $\alpha=0$, we will get a coexistence curve which is the same as the one obtained in Ref.\cite{Wei:2015iwa}. This can also be witnessed in Fig.\ref{1a}.

Actually, it is more interesting to study the nontrivial cases in which the effects of those parameters can not be ignored as $m\neq0$. Setting $\alpha=1$ (which means $\frac{12\sqrt{3}c_0c_1m^2\pi Q}{(k+c_0^2c_2m^2)^{3/2}V_2}=1$), from Eq.(\ref {cocur}) one can easily find out that the start point of the curve in the axis of reduced temperature will shift to $\sqrt{2}-1$ instead of $0$ when the reduced pressure approaches to zero. Indeed we can define a shifted temperature $\tau_s$ according to Eq.(\ref{cocur}) for convenience as
\begin{equation}
\tau_s=\frac{12\sqrt{3}c_0c_1m^2\pi Q}{12\sqrt{3}c_0c_1m^2\pi Q+\sqrt{2}(k+c_0^2c_2m^2)^{3/2}V_2}.
\end{equation}
From Fig.\ref{1b} we can see clearly the start point of the coexistence curve shifts to $(\tau_s,0)$ instead of $(0,0)$ in $p$--$\tau$ diagram, as if the original curve is suffering from a compressive deformation towards right.

\begin{figure*}
\centerline{\subfigure[]{\label{2a}
\includegraphics[width=8cm,height=6cm]{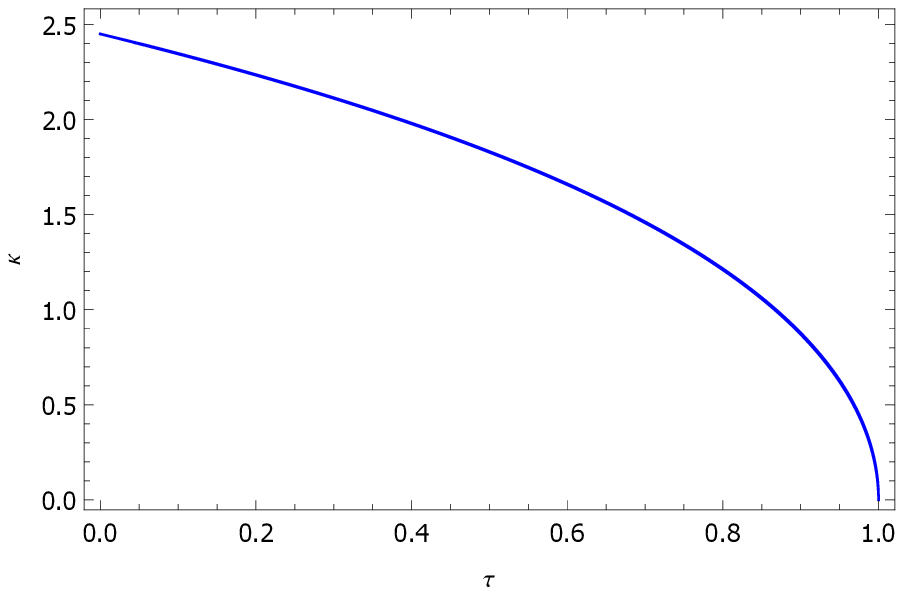}}
\subfigure[]{\label{2b}
\includegraphics[width=8cm,height=6cm]{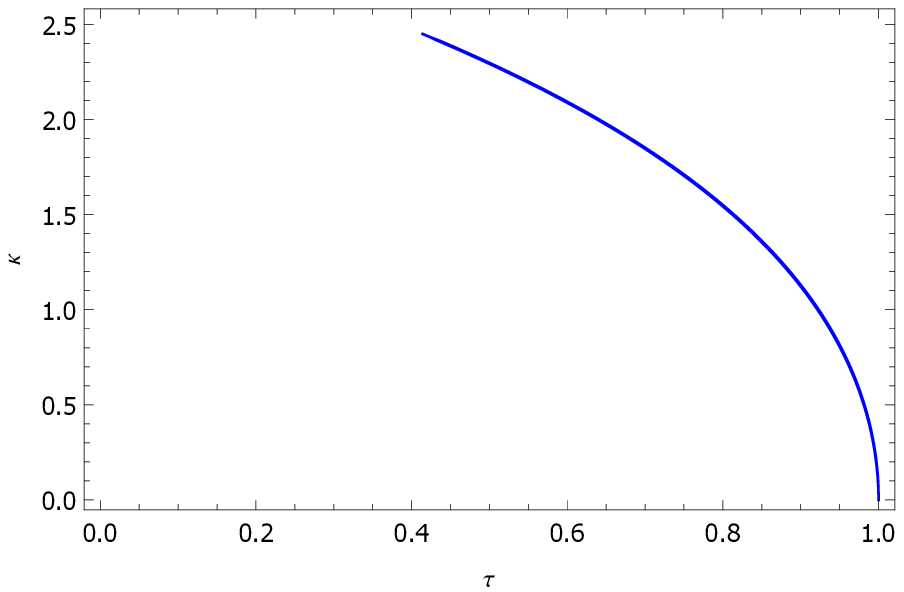}}}
 \caption{(a) the number density difference of the small and large massive gravity AdS black hole when $\alpha=0$ (b) the number density difference of the small and large massive gravity AdS black hole when $\alpha=1$} \label{fg2}
\end{figure*}

With the definition of number density $n=\frac{1}{v}$ for black hole molecules raised by Wei \emph{et al.}\cite{Wei:2015iwa}, one can obtain
\begin{equation}\label{numden}
\kappa\equiv\frac{n_1-n_2}{n_c}=\frac{\frac{1}{v_1}-\frac{1}{v_2}}{\frac{1}{v_c}}=\frac{\nu_2-\nu_1}{\nu_1\nu_2},
\end{equation}
where $\kappa$ depicts the difference of the black hole's molecule number densities between the small and large black hole. Here we have made the assumption that $\nu_2>\nu_1$.
Substituting the solutions of Eqs.(\ref{core1})-(\ref{core3}) into Eq.(\ref{numden}), we get
\begin{equation}
\kappa=\frac{n_1-n_2}{n_c}=\sqrt{6-6\sqrt{p(\tau)}},
\end{equation}
where $p(\tau)$ refers to Eq.(\ref{cocur}).
We will get exactly the same curve of the difference of the number densities for black hole molecules between the small and large black hole if we set $\alpha=0$. This can be witnessed in Fig.\ref{2a}. The nontrivial case of the $\kappa$ -- $\tau$ curve will also have the same compressive deformation as shown in Fig.\ref{2b} if $\alpha=1$.

\section{Conclusions and discussions}\label{Sec4}

We have explored the coexistence curve and the physical quantity describing the difference of the molecule number densities between the small and large $4$-dimensional AdS topological charged black hole in massive gravity. It is found that these results can reduce into forms which coincide with those obtained in general relativity RN-AdS black hole case if we ignore the parameters accociated with massive gravity, for instance, we may set the mass of graviton $m=0$. However, we will have a shifted temperature in the $p$--$\tau$ and $\kappa$ -- $\tau$ curves as $m\neq0$.

On the other hand, the horizon curvature $k$ can take non--positive value in massive gravity while there still exists a VdW--type phase transition. It is easy to check that the shifted temperature we defined will be zero when the mass of graviton can be ignored no matter we take the value of the horizon curvature $k$ to be $+1$ or $-1$. However, things get a little tricky when we set $k=0$. The shifted temperature will be $1$ once we take $k=0$, even we ignore the mass of graviton. How to explain this? Indeed, the condition of VdW--type phase transition is $k+c_0^2c_2m^2>0$\cite{Hendi:2017fxp}, so we can not set $k$ and $m$ both to be zero if we expect the phase transitions. Indeed, the shifted temperature will still be zero as long as we set the limit of $k$ to be zero but not equal to 0.

We not only have broadened the field of the knowledge concerning black hole phase transitions in massive gravity, but also have highlighted the rich physical meaning conveyed by the parameters in massive gravity.

\acknowledgments The authors acknowledge useful suggestions and discussions with Yong Zhang. This work is supported by the National Natural Science Foundation of China (Grant No. 11235003).

\end{document}